\documentclass[twocolumn,preprintnumbers,amsmath,amssymb]{revtex4-1}
\usepackage{graphicx}
\usepackage{ulem}
\usepackage[svgnames]{xcolor}
\usepackage{bm}
\usepackage{multirow}

\usepackage{float}
\usepackage{titlesec}
\usepackage{array}
\usepackage{booktabs}

\begin{document}

	\title{Quantum oscillations, Magnetic breakdown and thermal Hall effect in Co$_3$Sn$_2$S$_2$}
	
	\author{Linchao Ding$^{1,*}$, Jahyun Koo$^{2,*}$, Changjiang Yi$^{3,*}$, Liangcai Xu$^1$, Huakun Zuo$^{1}$,  Meng Yang$^{3}$, Youguo Shi$^{3}$,  Binghai Yan$^{2,\dag}$, Kamran Behnia$^{4, *}$, and Zengwei Zhu$^{1,\ddag}$}
	
	\affiliation{(1) Wuhan National High Magnetic Field Center and School of Physics, Huazhong University of Science and Technology,  Wuhan,  430074, China\\
		(2) Department of Condensed Matter Physics, Weizmann Institute of Science, 7610001 Rehovot, Israel\\
		(3)Beijing National Laboratory for Condensed Matter Physics, Institute of Physics, Chinese Academy of Sciences, Beijing, 100190, China\\
		(4)Laboratoire de Physique et Etude des Mat\'{e}riaux (CNRS/UPMC),Ecole Sup\'{e}rieure de Physique et de Chimie Industrielles, 10 Rue Vauquelin, 75005 Paris, France\\
	}
	
	\date{\today}
	
	\begin{abstract}
		
Co$_3$Sn$_2$S$_2$ is a ferromagnetic semi-metal with Weyl nodes in its band structure and a large  anomalous Hall effect below its Curie temperature of 177 K. We present a detailed study of its Fermi surface and examine the relevance of the anomalous transverse Wiedemann Franz law to it. We studied Shubnikov-de Haas oscillations along two orientations in single crystals with a mobility as high as $2.7\times$10$^3$ cm$^2$V$^{-1}$s$^{-1}$ subject to a magnetic field  as large as $\sim$ 60 T. The angle dependence of the frequencies is in agreement with density functional theory (DFT) calculations and reveals two types of hole pockets (H1, H2) and two types of electron pockets (E1, E2). An additional unexpected frequency emerges at high magnetic field. We attribute it to magnetic breakdown between the hole pocket H2 and the electron pocket E2, since it is close to the sum of the E2 and H2 fundamental frequencies. By measuring the anomalous thermal and electrical Hall conductivities, we quantified the anomalous transverse Lorenz ratio, which  is close to the Sommerfeld ratio ($L_0=\frac{\pi^2}{3}\frac{k_B^2}{e^2}$) below 100 K and deviates downwards at higher temperatures. This finite temperature deviation from the anomalous Wiedemann-Franz law is a source of information on the distance between the sources and sinks of the Berry curvature and the chemical potential.

		
			\end{abstract}
	\maketitle
	
	\section{Introduction}
Magnetic topological materials  have attracted tremendous recent attention\cite{PRL2014,Nakatsuji2015,Yan2017,Smej}. Co$_3$Sn$_2$S$_2$ is one of them, which  has a kagome lattice and becomes ferromagnetic below 177 K~\cite{Vaqueiro}. It is a half-metal, as the spin degeneracy of its carriers is almost entirely lifted in the ferromagnetic phase~\cite{Schnelle,Holder}. It is a semimetal with an equal number of electrons and holes. The density of these carriers is in the range of 10$^{19}$cm$^{-3}$~\cite{Ding2019}, so  that one electron  and one hole is shared by more than a thousand atoms.

The observation of a sizeable anomalous Hall effect in this context makes this solid an appealing platform. Because of the low carrier density, the anomalous Hall angle becomes remarkably large~\cite{LiuEnke2018,Lei2018}, as large as 33\% in iron-doped samples at low temperature \cite{AFM2020}. Angle-resolved photoemission spectroscopy~\cite{ChenYulin2019} and scanning tunneling microscopy~\cite{Morali2019} and magneto-optical studies~\cite{MO2020} confirm that is a magnetic Weyl semimetal with a topologically non-trivial electronic structure. The high mobility of carriers in this low-density semi-metal generates an ordinary Nernst coefficient~\cite{Behnia2016} on top of the anomalous Nernst response. A study of its thermoelectric response~\cite{Ding2019} has documented the contrasting evolution of the ordinary and anomalous  components of the  Nernst response. As expected in a Berry-curvature picture, increasing disorder enhances the anomalous Nernst effect and reduces the ordinary Nernst effect.~\cite{Ding2019}.
	
In this paper, we present two distinct experimental studies of Co$_3$Sn$_2$S$_2$. First of all, we report on the fermiology of this system by  studying quantum oscillations on high-mobility crystals in presence of strong magnetic fields.  Our sample has a  residual resistivity ratio (RRR) of 30 and its mobility is as large as 2.7$\times$10$^3$ cm$^2$V$^{-1}$s$^{-1}$. Using pulsed magnetic fields up to 61 T and low temperatures, we tracked quantum oscillations with magnetic field perpendicular and parallel to the $z$-axis([0001]). The angle dependence of the observed frequencies was compared with what is expected according to the density functional theory (DFT) calculations. There were consistent. In addition, we found an unexpected frequency emerging at high field and attributed it to a magnetic breakdown orbit between a hole-like and an electron-like orbit.  In a type II Weyl semimetal, magnetic breakdown between Weyl nodes in momentum space is reminiscent of Klein tunneling ~\cite{Brien2016}. However, the observed breakdown frequency is close to the sum of the two fundamental frequencies (and not their difference, which was predicted~\cite{Brien2016}). In this respect, magnetic breakdown in Co$_3$Sn$_2$S$_2$ is reminiscent of what has been observed in WTe$_2$~\cite{Zhu2015} and in  elemental metals, Mg~\cite{Shoenberg}  and Be~\cite{Shoenberg} and contrasts with what was reported in the case of ZrSiS\cite{Muller2020}. The second component of the present report is a quantitative study of the thermal Hall effect. This complements previous studies of the anomalous Hall~\cite{LiuEnke2018,Lei2018,Ding2019} and anomalous Nernst~\cite{Ding2019} effects in this system. Putting the anomalous Lorenz ratio (i.e. the ratio of anomalous thermal and electrical conductivities divided by temperature) under scrutiny, we found that the anomalous transverse Wiedemamn-Franz law ((WF law)) holds at low temperature, but a downward deviation starts above 100K. In this regard, Co$_3$Sn$_2$S$_2$ is reminiscent of Mn$_3$Ge~\cite{Xu2020} and contrasts with  Mn$_3$Sn~\cite{Li2018}, in which the anomalous WF law holds even at room temperature. As argued previously~\cite{Xu2020}, the temperature at which the departure from the anomalous WF law occurs depends on the details of the Berry curvature in the vicinity of the Fermi level.
	
\section{Experimental and computational details}

High quality single crystals of Co$_3$Sn$_2$S$_2$ were grown by the self-flux method as in a previous work \cite{Growth2020}. A stoichiometric ratio of Co, Sn and S powders was sealed in a quartz tube. The quartz tube was loaded into a box furnace and heated to 673 K and held for 24 hours. Then the tube was heated to 1273 K and held for 12 hours, then slowly cooled to 1073 K and the furnace was turned off. Shining crystals were obtained by mechanical separation carefully. The low-field transport measured by a standard four-probe method was performed on a physical property measurement system (Quantum Design PPMS-9). High-field magnetotransport measurements were carried out under pulsed magnetic field in Wuhan National High Magnetic Field Center, each contact resistance was less than 3 $\Omega$ in all the measurements.

Our calculations were performed using the density-functional theory (DFT) in the framework of the generalized gradient approximation \cite{PBE1996}) with the Vienna \textit{ab-intio} package\cite{Kresse1999}. To interpolate the 3D Fermi surfaces, we obtained by a tight-binding Hamiltonian, based on localized Wannier functions \cite{Mostofi2008} projected from the DFT Bloch wave functions. We note that the orientations of $x$ is defined as the current direction, and the angle between $x$ and the [2\={1}\={1}0] orientation is about 10$^\circ$ (see more details in supplementary material \cite{SM} and Fig.\ref{basicinfo}a), $z$ is defined as [0001] orientation.

	\begin{figure}
		\includegraphics[width=9.5cm]{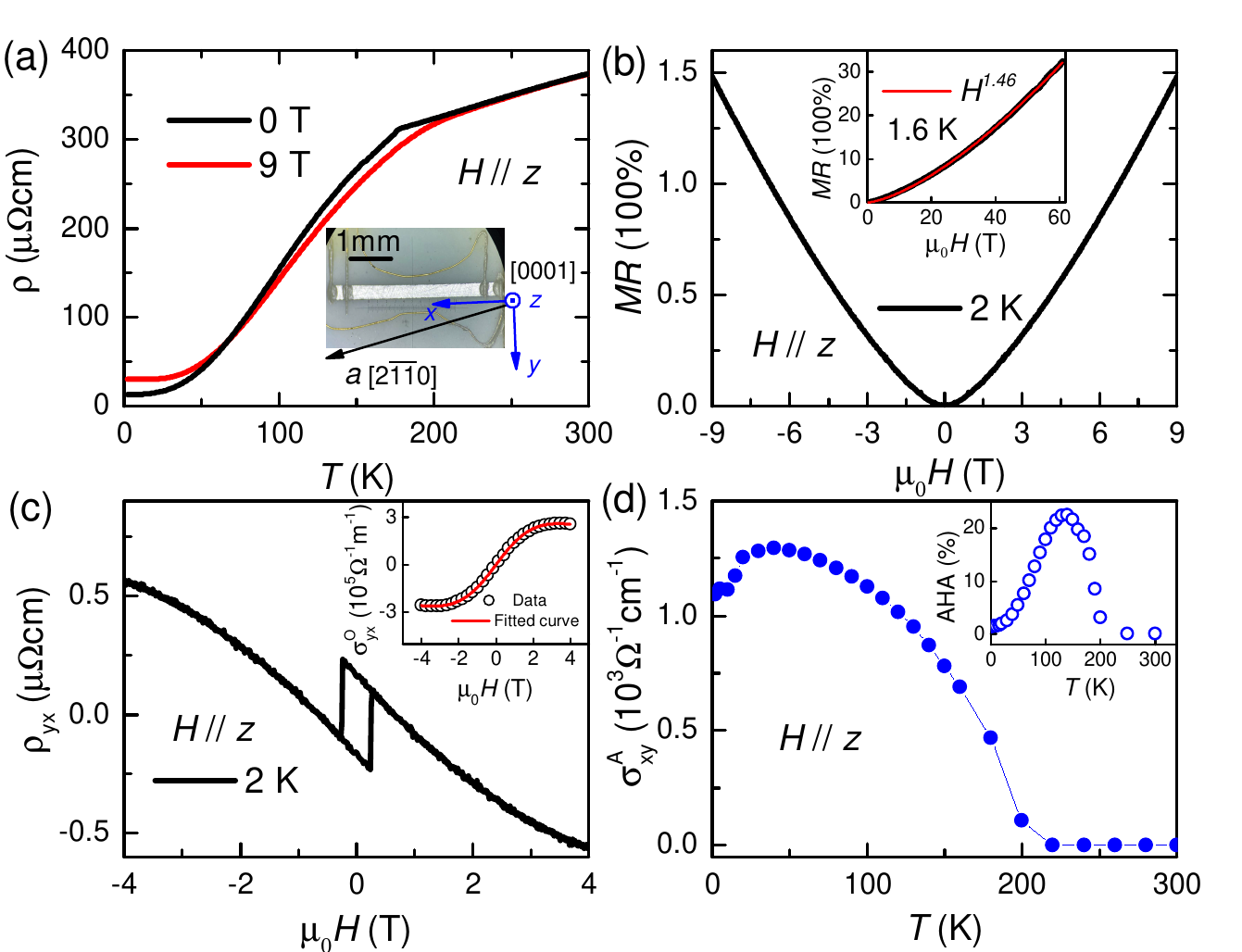}
		\caption{ 
			(a) Temperature dependence of resistivity at zero-field (black) and 9 T (red) along $z$-axis. (b) Magnetoresistance measured in fields up to 9 T at 2 K, showing a non-saturated positive magnetoresistance, the inset shows the  non-saturating magnetoresistivity up to 61 T at 1.6 K. (c) Hall resistivity at 2 K. The inset compares the normal part of Hall conductivity (black hollow circle) with  a two-band fit (red line). (d) Temperature dependence of the anomalous Hall conductivity and the the anomalous Hall angle (in the inset). }
		\label{basicinfo}
	\end{figure}
	\begin{figure*}
	\includegraphics[width=18cm]{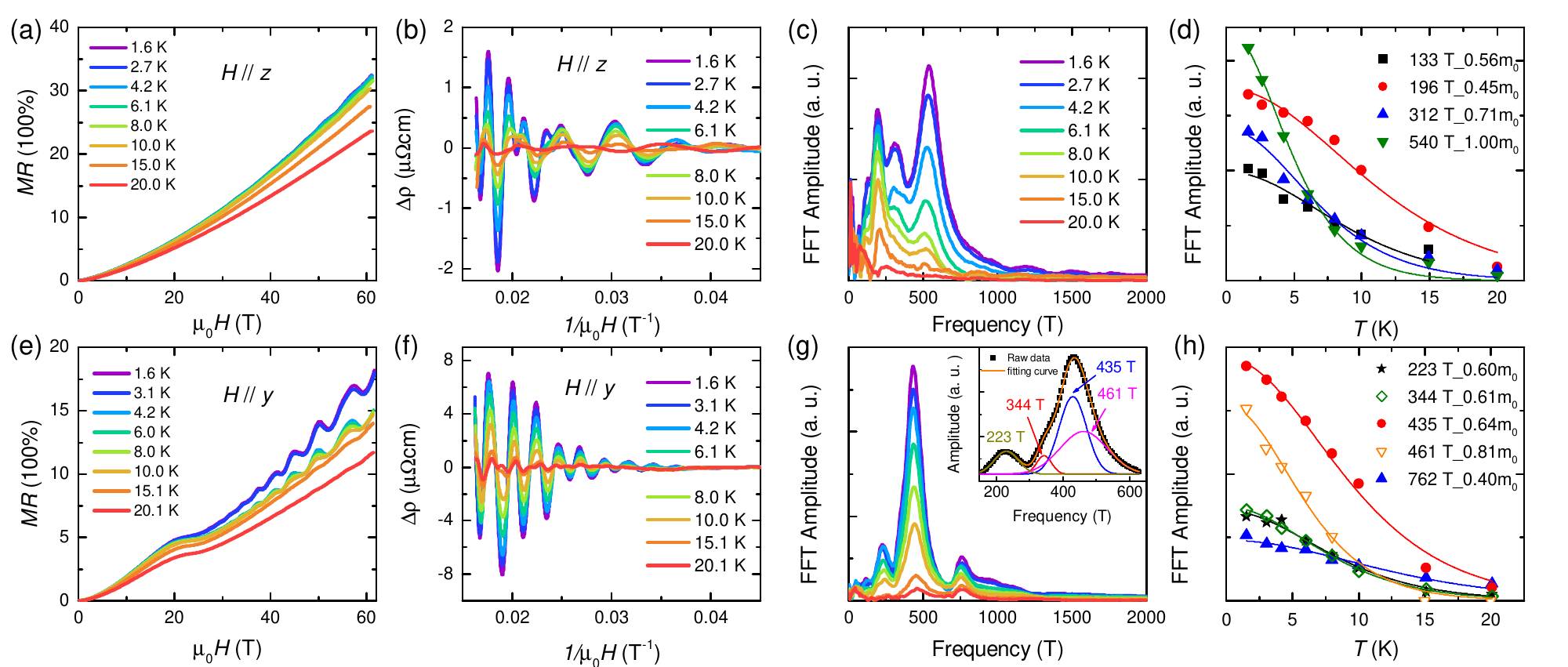}
		\caption{ (a) and (e) the magnetoresistance as a function of the magnetic field $\mu_0H$ at different temperatures for $H\parallel z-$axis and $H\parallel y-$axis respectively. (b) and (f) The oscillatory part of resistivity $\Delta\rho$ for $H\parallel z-$axis and $H\parallel y-$ respectively. (c) and (g) show the fast Fourier transformation spectrum at various temperature. The inset shows the multi-peak fitting used to distinguish between superposed frequencies. (d) and (h) the L-K fitting with the effective masses}
		\label{SdH}
	\end{figure*}
	
	\section{Carrier density, mobility and the anomalous Hall conductivity}

Fig. \ref{basicinfo}(a) shows the temperature dependence of resistivity measured as the current was applied along $x$-axis at zero field (black) and 9 T (red) along $z$-axis. A kink at Curie temperature $\sim$175 K is visible in $\rho (T)$ at zero field. The temperature dependence of resistivity shows a metallic behavior with a residual resistivity rate of RRR $=\rho (300\rm{K})/\rho (2\rm{K})\sim 30$. This is larger than what has been reported previously~\cite{LiuEnke2018, Lei2018, Ding2019, Tanaka2020,nano2020}, indicating the high quality of the sample. Accordingly, the residual resistivity $\rho_{0} = 12 \mu\Omega$cm is also smaller than was found before  \cite{LiuEnke2018,Lei2018}. The higher mobility leads to a larger magnetoresistance. This  has been explored  in other semi-metals, such as WTe$_2$ \cite{Ali2014,Zhu2015}, WP$_2$ \cite{Kumar2017}, PtBi$_2$  \cite{Gao2017,Zhao2018} and in Sb \cite{Fauque2018}. Fig. \ref{basicinfo}(b) displays the non-saturated positive magnetoresistance (MR = $ [\rho_{xx}(\mu_0H)-\rho_{xx}(0)]/\rho_{xx}(0) \times 100\%$) to 150\% under the magnet field of 9 T. As shown in the inset, the MR curve shows $H^{1.46}$ dependence up to 61 T at 1.6 K.

Fig. \ref{basicinfo}(c) shows the nonlinear field dependence of the Hall resistivity at 2 K. The inset shows the field dependence of the ordinary Hall conductivity  ($\sigma_{xy} = \rho_{yx}/(\rho_{yx}^2+\rho_{xx}^2)$) after subtracting anomalous part\cite{Ding2019}. The red line represents a two-band fit to the data, using the expression:

	\begin{equation}
		\sigma_{yx}^{O} = [\frac{\color{black}p\color{black}\mu_h^2}{1 + \mu^2_{h}(\mu_0H)^2}-\frac{\color{black}n\color{black}\mu_e^2}{1 + \mu^2_{e}(\mu_0H)^2}]e(\mu_0H)
	\end{equation}

This allowed us to deduce the carrier density of holes (electrons): $p=8.9\times 10^{19}$ cm$^{-3}$ ($n=8.7\times 10^{19}$ cm$^{-3}$) and hole (electron) mobilities of $\mu_h=2713$ cm$^2$V$^{-1}$s$^{-1}$ ($\mu_e=2673$ cm$^2$V$^{-1}$s$^{-1}$). The average transport mobility  can also be extracted from residual resitivity at zero magnetic field
 $\rho_{0}$, using $\mu_0=\frac{1}{\rho_{0}e(n+p)} = 2969$ cm$^2$V$^{-1}$s$^{-1}$.

Fig. \ref{basicinfo}(d) shows the temperature dependence of the anomalous Hall conductivity (AHC) attains 1290 $\Omega^{-1}$cm$^{-1}$ at 40 K and 1093 $\Omega^{-1}$cm$^{-1}$ at 2 K, the anomalous Hall angle AHA$=\sigma_{xy}/\sigma_{xx}\times 100\%$ reaches $23\%$ at 140 K, similar to previous reports\cite{LiuEnke2018,Lei2018}. In contrast to dirtier samples, the anomalous Hall conductivity  slightly decreases below  50 K. Such a behavior  was previously noticed in cleaner samples~\cite{Ding2019}.

			\begin{figure*}
		\includegraphics[width=16cm]{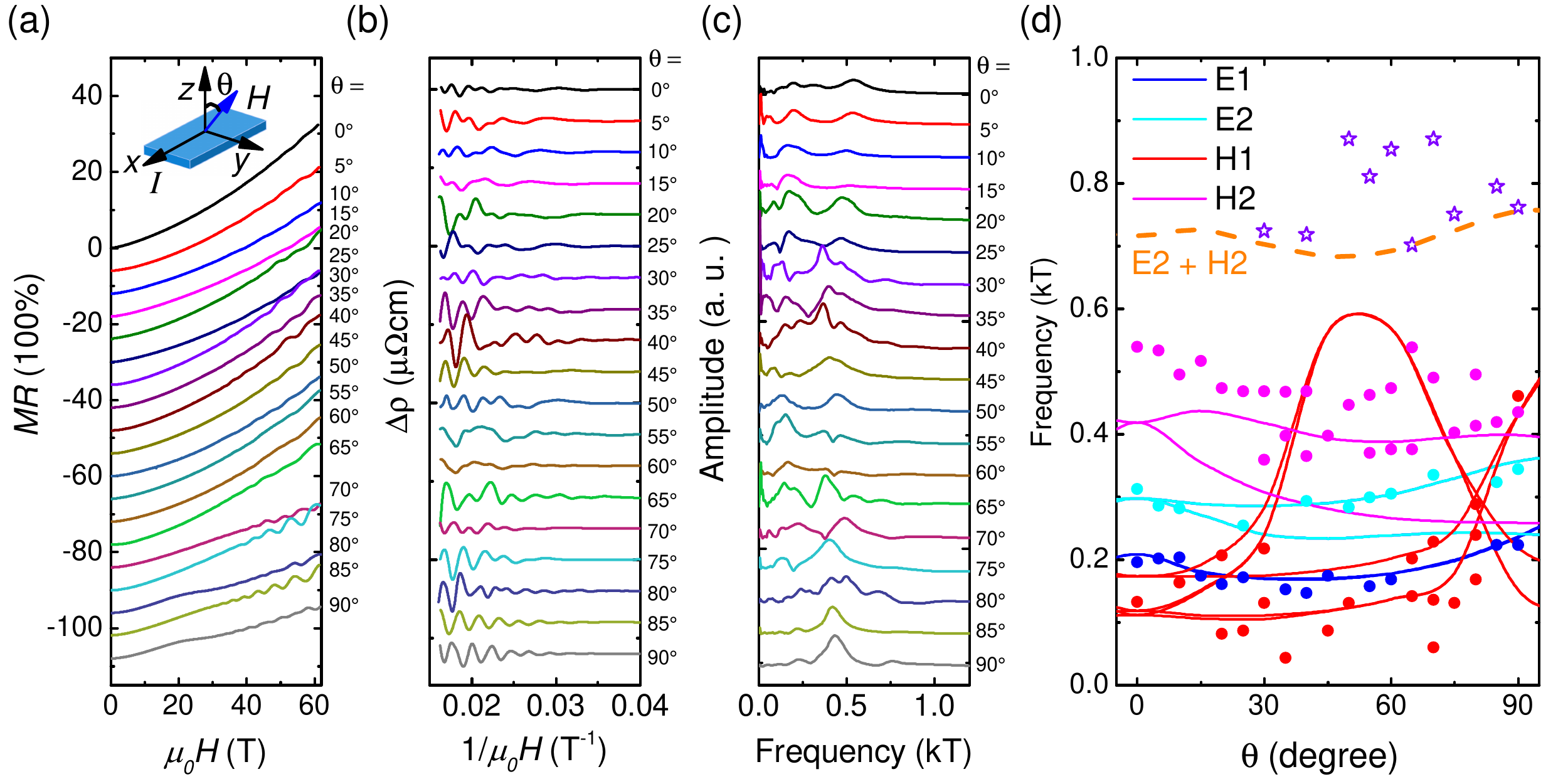}
		\caption{ (a) Field dependence of MR at 1.6 K with the field rotating from the $z-$axis to the $y-$ axis. The field is always perpendicular to the current direction. (b) SdH oscillatory component as a function of 1/$\mu_0H$ at different angles. (c)FFT spectra of SdH at various angles. (d) Angular dependence of the frequencies extracted from the SdH oscillations, the lines are from the theory calculation results.}
		\label{Angle-SdH}
	\end{figure*}

\section{The Fermi surface}
	
Fig. \ref{SdH} (a) and Fig. \ref{SdH} (e) display the temperature dependence of magnetoresistance under high magnetic field along $z$-axis and $y$-axis. Quantum oscillations of resitivity, known as the Shubnikov-de Haas (SdH) effect,  are observed when the magnetic field exceeds 20 T. A plateau in magnetoresistance appears around 20 T when the magnetic field is parallel to  the $y$-axis. When the magnetic field is parallel to the $xy$ plane, the saturation field for magnetization is about 23 T ~\cite{Shen2019}. Therefore, the resistivity plateau is concomitant with the field-induced rotation of the magnetic moments from $z$-axis to the $xy$-plane.

As shown in Fig. \ref{SdH} (b) and (f), the SdH oscillations become clearly visible by subtracting a polynomial background from the data. Fig. \ref{SdH} (c) and (g) present the fast Fourier transformation (FFT) spectrum of the oscillating component, $\Delta \rho$ at different temperatures. Some of the broad asymmetric peaks in  the FFT spectrum , such as the peak around 430 T for $H\parallel y$  are caused by superposition of multiple peaks.  We fit them, using  the Gauss multi-peak method\cite{YBCO2009}. The evolution of these peaks at different temperatures is shown in the inset of the Fig.2g.

We can identify four peaks in the case of $H\parallel$ $z$ and five peaks in $H \parallel y$, respectively. These frequencies observed in FFT spectra are closely related with the orthogonal extremal cross-sectional area ($A_F$) of the Fermi pockets by the Onsager relation $F = ( \hbar/2\pi e) A_F $. We extracted the effective mass by fitting the temperature dependence of the oscillation amplitude with the thermal damping term $R_T=X/\rm{sinh}(x)$ of the Lifshitz-Kosevich (L-K) formula \cite{LK1956}, where $X=14.69m^*T/B$ and $m^*$ is the effective cyclotron mass. We obtain the effective masses of different frequencies as shown in the Fig. \ref{SdH}(d) and (h).  The parameters for pockets in two orientations are summarized in Table \ref{Table:SdH}. We note that the Dingle mobility is about 30 times lower than the transport mobility. A much larger difference between the two mobilities has been reported in other semi-metals, such as Sb, WP$_2$ and Cd$_3$As$_2$. In all cases, transport mobility  is significantly larger than quantum mobility. In other words, the mean-free-path for momentum relaxation is longer than the one associated with the distance between collisions which broaden the Landau levels~\cite{Jaoui2020}.

	\begin{table*}[!htb]
		\caption{	The derived parameters from SdH oscillations for Co$_3$Sn$_2$S$_2$. $F$ is the quantum oscillation frequency, $A_{F}$ is the extremal cross-sectional area of the pockets in unit of $10^{-3}$\AA$^{-2}$, $m^{*}$ is the effective mass (in terms of $m_e$) and the Fermi energy $E_F$ for each pocket, $\mu_q$ is quantum mobility in unit of cm$^2$/Vs.
		}
		\begin{tabular}{
				c|c c c c| c cccccc}
			
			\hline
			\hline
			
     \rule{0pt}{9pt}	$direction$  &  & $H$$\parallel$$z$  &  &    &  & &$H$$\parallel$$y$&   \\
			\hline
			
			\rule{0pt}{9pt}	$label$ & E1 & E2 & H1 & H2 & E1& H1&  E2 & H2 & H1'     \\
			\hline
			\rule{0pt}{10pt} $F_{theory}$ & 208 & 297 & 118&418&235 &219  & 356  &  395  & 432 \\
			
			\rule{0pt}{10pt} $A_F(theory)$($10^{-3}$\AA$^{-2}$) & 19.9 & 28.4 &11.2 &39.9 & 22.4& 20.9  & 34  &  37.7  & 41.8 \\
			
		\hline
		
			\rule{0pt}{10pt} $F_{exp.}$ (T) & 196 & 312  & 133& 540& & 214  & 343  & 429   & 462 \\
			
			\rule{0pt}{9pt}	$A_F$ ($10^{-3}$\AA$^{-2}$) &18.9&30.3&12.7&51.5& &20.4&32.7&40.9&44.1 \\
			\rule{0pt}{9pt}	$m^{*}$ (m$_e$)& 0.45&0.71 &0.56 &1.00&&0.52&0.64&0.53&0.81&\\
			
			\rule{0pt}{9pt}	$E_F$ (meV)   &50.4 &50.8&27.5 &62.5 &&47.6&62.0&93.7&66.0 \\
			
			\rule{0pt}{9pt}	$T_D$ (K)    & 30.8 & && & &15.1& 22.9 &  & &  \\
			
          	\rule{0pt}{9pt}		$\mu_q$(cm$^2$/Vs)   & 106 & & & & & 221& 191 & &&  \color{black}	 \\

			\hline
			\hline
		\end{tabular}
		
		\label{Table:SdH}
	\end{table*}

We studied the evolution of the SdH oscillations  at 1.6 K with  rotation. Fig. \ref{Angle-SdH}(a) presents the angle-dependence of MR which decreases gradually when the field rotates away from the $z$-axis to the $y$-axis. $\theta$ is defined by the angle between the field and the $z$-axis, the MR reaches a minimum value at $\theta = 90^{\circ}$ when the field is parallel to the $y$ axis. Fig. \ref{Angle-SdH}(b) shows SdH oscillations for each angle after subtracting a fitted polynomial. The angle dependencies of the frequencies extracted from the SdH oscillations are summarized in Fig. \ref{Angle-SdH} (d). Symbols  show the data and lines represent the DFT calculations.

\begin{figure}
		\includegraphics[width=9cm]{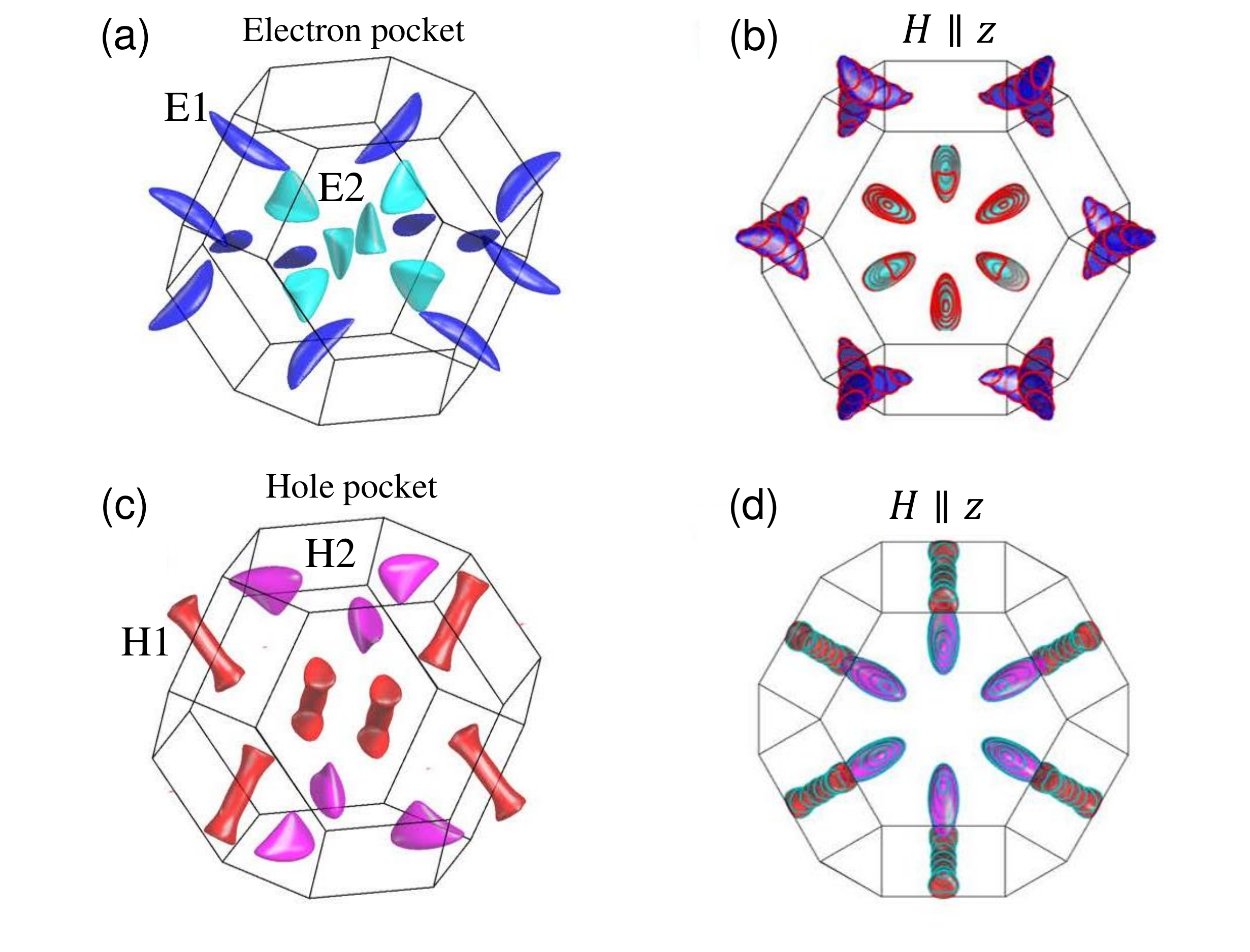}
		\caption{The calculated Fermi surface of Co$_3$Sn$_2$S$_2$. (a) The electron pocket with two types, banana-shaped pockets called E1 with blue color and triangular shape pockets called E2 with cyan color. (c) The hole pocket with two types, cylinder-shaped pockets called H1 with a red color and triangular shape pockets called H2 with a purple color. The possible cross-section with magnetic field align $z$ direction for the electron pocket with a red line (b) and for the hole pocket with blue line (d).
		}
		\label{Fig: FermiPocket}
	\end{figure}

According to the DFT calculations, Co$_3$Sn$_2$S$_2$ has two types of electron pockets and two types of hole pockets. We labeled each pocket as shown in Fig. \ref{Fig: FermiPocket}. In the case of electrons, there are twelve banana-shape pockets(E1) across the Brillouin zone, and six triangle-like pockets (E2) near the $\Gamma$ point shown in the Fig. \ref{Fig: FermiPocket}(a). In the case of holes, there are six cylinder-like pockets (H1) across the Brillouin zone and six triangular-like pocket(H2) close to the k$_z$ boundary shown in the Fig. \ref{Fig: FermiPocket}(c).

We rotated the magnetic field in the $yz$ plane.  When the magnetic field is aligned along the $z$-axis ($\theta=0$) the extreme cross-section is identical for each type of pocket and one expects to see four distinct frequencies. When the magnetic field has a finite angle with the axis, the extreme cross-section degeneracy of pockets of the same type is lifted and more than four frequencies is expected. The calculated H1 pocket can have two maximum extreme cross-section and one local minimum cross-section, since it has a quasi- cylindrical shape. Note that the calculated SdH frequencies are not symmetric between +$z$ to $y$ direction and --$z$ to $y$ direction.  (see the SM for more discussion\cite{SM}).
	
Fig. \ref{Fig: FermiPocket} shows the calculated Fermi surface pockets and their projections along high-symmetry $z$-axis. The electron pockets are in blue and the hole pockets are in red. The Fermi pockets distribution follow the  M$_y$ symmetry and inversion symmetry. The total volume of the Fermi pockets  yields the the theoretical carrier densities. The calculated hole density is 1.82 $\times$ 10$^{20}$ cm$^{-3}$ and for the electron carrier density is 1.59 $\times$ 10$^{20}$ cm$^{-3}$ at $\mu = -2 $ meV with respect to the charge neutral point. As discussed above,  our experimental Hall conductivity together with two-band fitting yields a carrier density, which is approximately twice lower. A similar discrepancy between experimental and theoretical carrier density was observed in the case of WTe$_2$~\cite{Zhu2015}.

\begin{figure}
	\includegraphics[width=9cm]{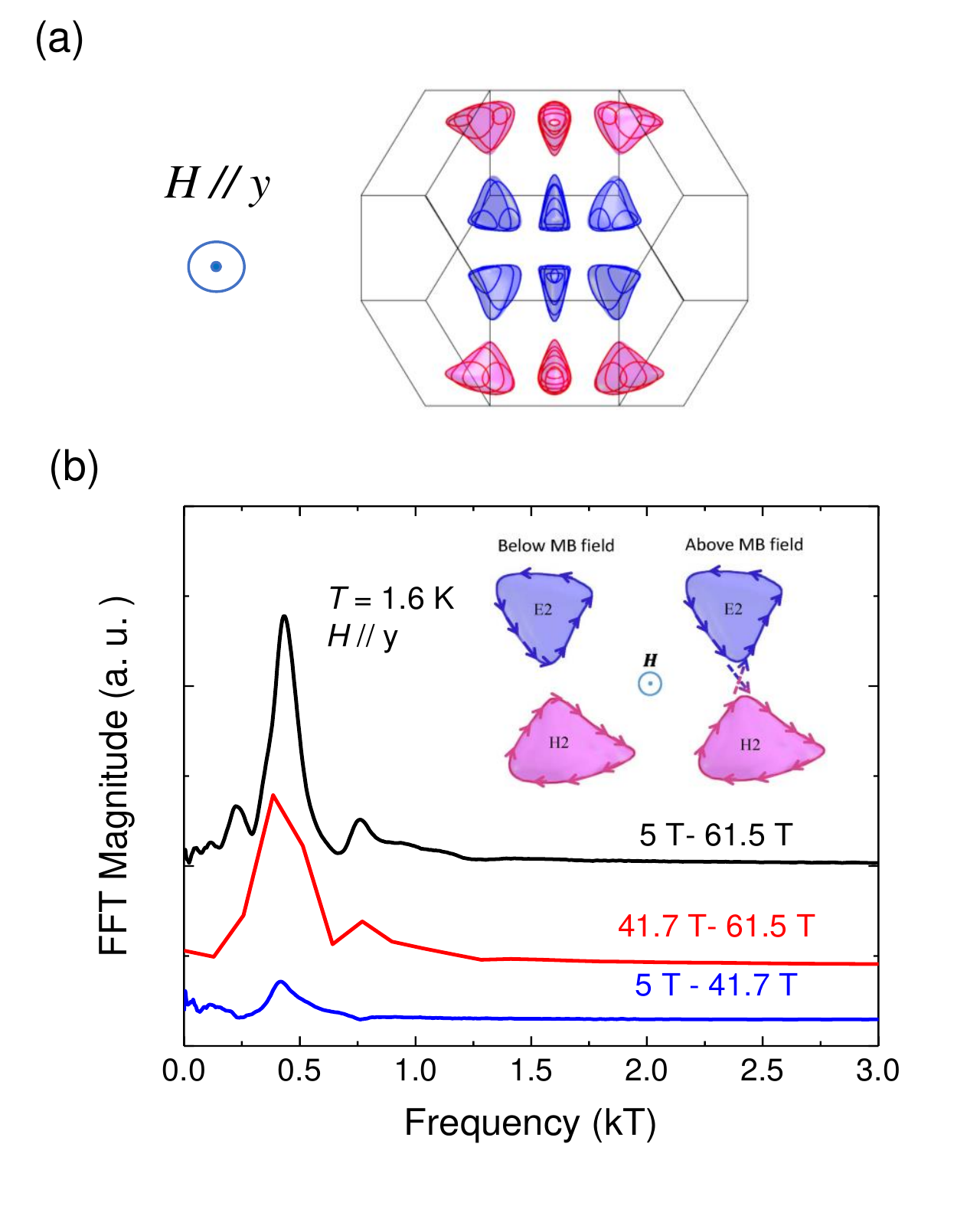}
	\caption{(a) Projections of the E2 and H2 pockets inside the Brillouin zone for $H\parallel y$.  The pockets are close to each other in the momentum space allowing tunneling at high field. (b)FFT analysis of the SdH oscillations at $\theta=90^{\circ}$ in  different field windows. For fields lower than 41.7 T (black line), the 760 T peak is absent. With increasing magnetic field, the peak associated with  this frequency increases in amplitude. The inset is a sketch of the semi-classical magnetic breakdown (MB).}
	\label{FFT}
\end{figure}

\section{Magnetic breakdown}

As shown in the Fig. \ref{Angle-SdH}(d), there is a satisfactory consistency between experimental and theoretical results. DFT calculations do not expect any  frequency larger than 600 T. However, a FFT peak is present at 760 T, when the magnetic field is parallel to the $y$-axis. We attribute this frequency to a magnetic breakdown orbit. As shown in Fig. \ref{Angle-SdH}(d),  this frequency is observed in the angular range of  $\theta = 30^{\circ}$ to  $\theta = 90^{\circ}$  and it corresponds roughly to sum of E2 and H2 frequencies. Fig.\ref{FFT}(a) shows the cross sections of the E2 and H2 pockets when the field is parallel to $y$-axis.

For $\theta = 90^{\circ}$, the 760 T frequency suddenly appears above a threshold field of 41.7 T. This can be seen  in Fig. \ref{FFT}(b), which presents the FFT spectrum for different field windows. We suggest that above 41.7 T  magnetic breakdown occurs between the electron pocket E2  and the hole pocket H2, which according to the DFT calculations are close to each other in the momentum space.  When the field is in the $y$ axis ($\theta = 90^{\circ}$), each of these pockets has two distinct extreme orbits and therefore two distinct frequencies.  The  fundamental frequencies of 343 T (for E2) and  429  (for H2) are the larger ones and their sum is close  to the observed frequency of 760 T. The angle dependence of the magnetic breakdown frequency is roughly consistent with the sum of fundamental frequencies of the E2 and H2 in the Fig.\ref{Angle-SdH}(d). The breakdown field which is defined as $H_B$ varies with rotation between 35 T and 48 T (See the SM for more details).

Magnetic breakdown in momentum space between an electron and hole pocket resembles the Klein tunneling at a $p-n$ junction in real space~\cite{Brien2016}.  We note, however, that theory expects that this Klein tunneling leads to a breakdown frequency equal to the difference between the two original fundamental frequencies and not the sum as we observe here. Previous experiments have reported the observation of magnetic breakdown  in two other topological materials, WTe$_2$\cite{Zhu2015} and ZrSiS\cite{Muller2020}. In the case of WTe$_2$, the breakdown  frequency  is the sum of two fundamental frequencies. But, this breakdown may occur between the adjacent two hole pockets.  In ZrSiS\cite{Muller2020}, magnetic breakdown is found to happen between pockets of opposite signs. There are multiple breakdown frequencies, which include not only the difference between frequencies, but also different combinations fundamental frequencies.

		\begin{figure}
		\includegraphics[width=8cm]{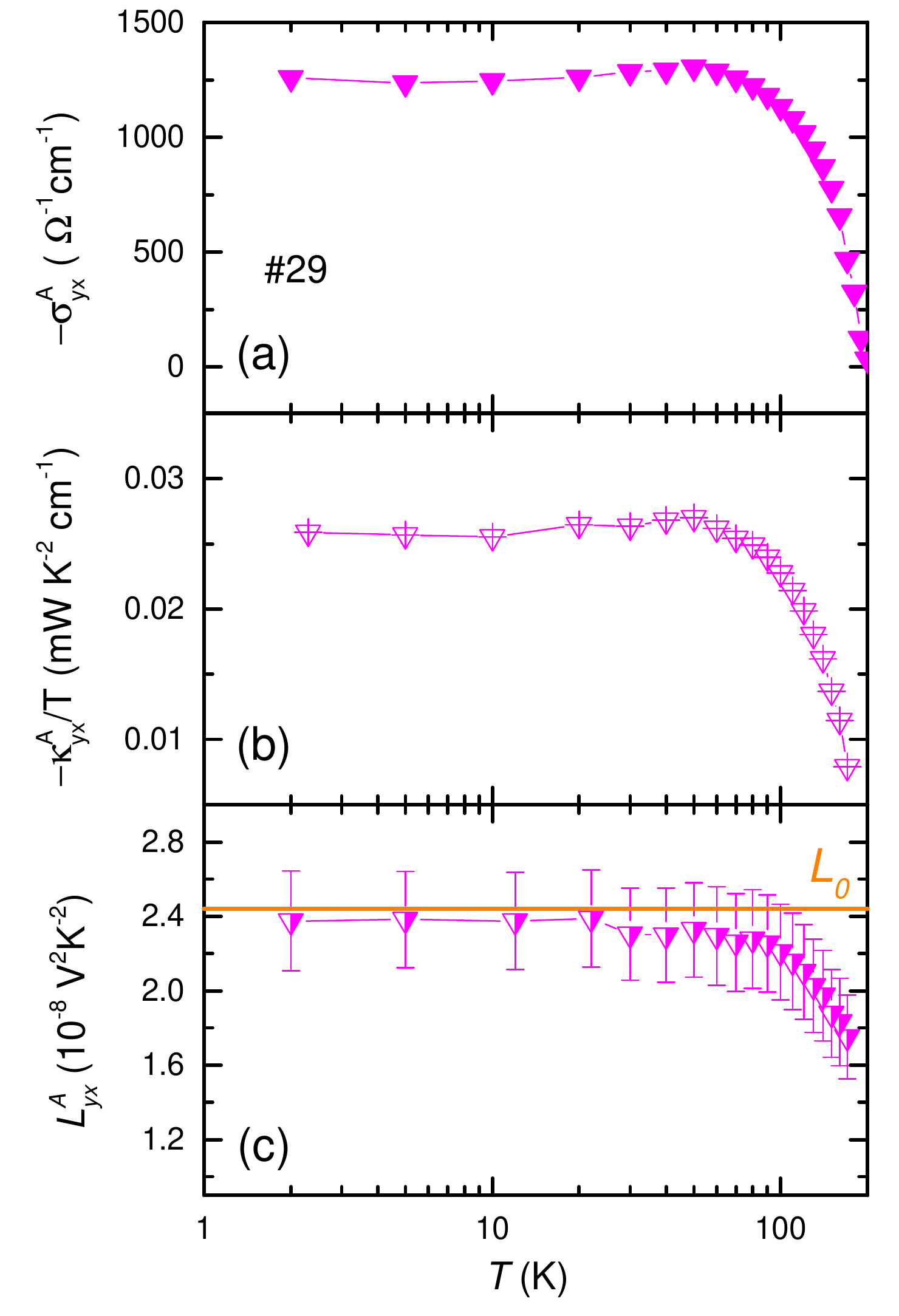}
		\caption{The temperature dependence of the anomalous Hall conductivity (a), the anomalous thermal Hall conductivity (b)  and the anomalous Lorenz number (c).  The anomalous WF law is obeyed at low temperature, but a deviation starts above 100 K. }
		\label{Fig: WF}
	\end{figure}
	
\section{The anomalous thermal Hall effect and the Wiedemann-Franz law}

Non-vanishing summation of the electronic Berry curvature is known to generate the intrinsic anomalous Hall effect (AHE). Measuring the thermoelectric (Nernst)  and the thermal (Righi-Leduc) counterparts of the anomalous Hall effect provides additional information on the topological origin of the AHE. The amplitude of the anomalous transverse thermoelectric conductivity has been found to correlate with the amplitude of the anomalous Hall conductivity\cite{Xu2020b}. In this section we focus on the correlation between anomalous electrical and thermal Hall conductivities.

Fig.\ref{Fig: WF}(a,b) show the temperature dependence of the anomalous Hall conductivity ($-\sigma_{yx}^A$) and the anomalous thermal Hall conductivity ($-\kappa_{yx}^A/T$) in another sample \#29 (See the supplement for the raw data\cite{SM}), grown by chemical vapor transport method \cite{Ding2019}. Below the Curie temperature, both  increase rapidly with decreasing temperature and saturate to an almost temperature-independent magnitude below 100 K. The magnitude of $\sigma_{yx}^A$ is  consistent with the previous reports\cite{LiuEnke2018,Lei2018, Ding2019}. Our $\kappa_{yx}^A/T$ data allows to check the validity of the anomalous Wiedemann-Franz Law for transverse response \cite{Li2018,Xu2020} by comparing the anomalous Lorenz number $L_{yx}^A=\frac{\kappa_{yx}^A}{\sigma_{yx}^AT}$ with the Sommerfeld value $L_0=\frac{\pi^2}{3}(\frac{k_B}{e})^2$ where $e$ and $k_B$ are the elementary charge of electron and the Boltzmann constant, respectively.

As seen in Fig.\ref{Fig: WF}(c), the Wiedemann-Franz Law holds below 100 K. This confirms once again that the anomalous Hall response is a Fermi surface (and not a Fermi sea) property as originally argued by Haldane \cite{Haldane}. However, above 100 K, a downward deviation in the the Wiedemann-Franz Law emerges. As argued previously \cite{Xu2020}, this deviation can be understood as the manifestation of the way the Berry curvature is averaged by thermal and electric probes. In a manner similar to Mn$_3$Ge \cite{Xu2020}, the downward deviation from the WF law in  Co$_3$Sn$_2$S$_2$ arises when  $\sigma_{yx}^A$ is evolving with temperature. Interestingly, according to a recent study of optical conductivity, in the ferromagnetic phase of Co$_3$Sn$_2$S$_2$, bands shift with increasing magnetization on Co sites. Such an evolution with temperature approaching the Curie temperature can generate a difference between the way Berry curvature is integrated in  the electrical and the thermal channels \cite{Yang2020}.

\section{Summary}
In summary, we performed a study of quantum oscillations of magnetoresistance in high-quality single crystals of Co$_3$Sn$_2$S$_2$ up to 61 T. We found a good agreement between the frequency of these oscillations and the Fermi surface of Co$_3$Sn$_2$S$_2$ according to DFT calculations. The Fermi surface consists of two types of electron pockets and two types of hole pockets. We detected an additional frequency and attributed it to magnetic breakdown between an electron and a hole pocket. We also found that the anomalous transverse Wiedemann-Franz law holds at low temperature, but a downward deviation emerges at finite temperature.

\section{Acknowledgments}
This work was supported by the National Science Foundation of China (Grant No. 51861135104 and 11574097), the National Key Research and Development Program of China (Grant No.2016YFA0401704),  and the Fundamental Research Funds for the Central Universities (Grant no. 2019kfyXMBZ071). B.Y. acknowledges the financial support by the Willner Family Leadership Institute for the Weizmann Institute of Science, the Benoziyo Endowment Fund for the Advancement of Science, Ruth and Herman Albert Scholars Program for New Scientists, and the European Research Council (ERC) under the European Union's Horizon 2020 research and innovation programme (grant agreement No. 815869). K. B. was supported by the Agence Nationale de la Recherche (ANR-18-CE92-0020-01; ANR-19-CE30-0014-04).

	\noindent\
	\dag  \verb|binghai.yan@weizmann.ac.il|\\
	*\verb|kamran.behnia@espci.fr|\\
	\ddag  \verb|zengwei.zhu@hust.edu.cn|\\

	\clearpage
	
	\renewcommand{\thesection}{S\arabic{section}}
	\renewcommand{\thetable}{S\arabic{table}}
	\renewcommand{\thefigure}{S\arabic{figure}}
	\renewcommand{\theequation}{S\arabic{equation}}
	
	\setcounter{section}{0}
	\setcounter{figure}{0}
	\setcounter{table}{0}
	\setcounter{equation}{0}
	
	{\large\bf Supplemental Materials for "Bulk Fermi surface of the ferromagnetic Kagome-lattice semimetal Co$_3$Sn$_2$S$_2$"}

	\begin{figure}
		\includegraphics[width=7cm]{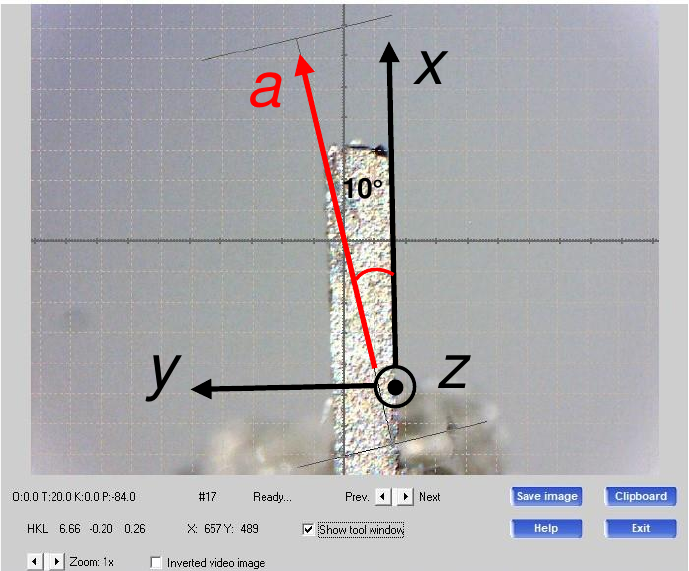}
		\caption{ The in-plane orientation of the sample was determined by a single crystal XRD.}
		\label{SM.xyz}
	\end{figure}

	\begin{figure}
		\includegraphics[width=9cm]{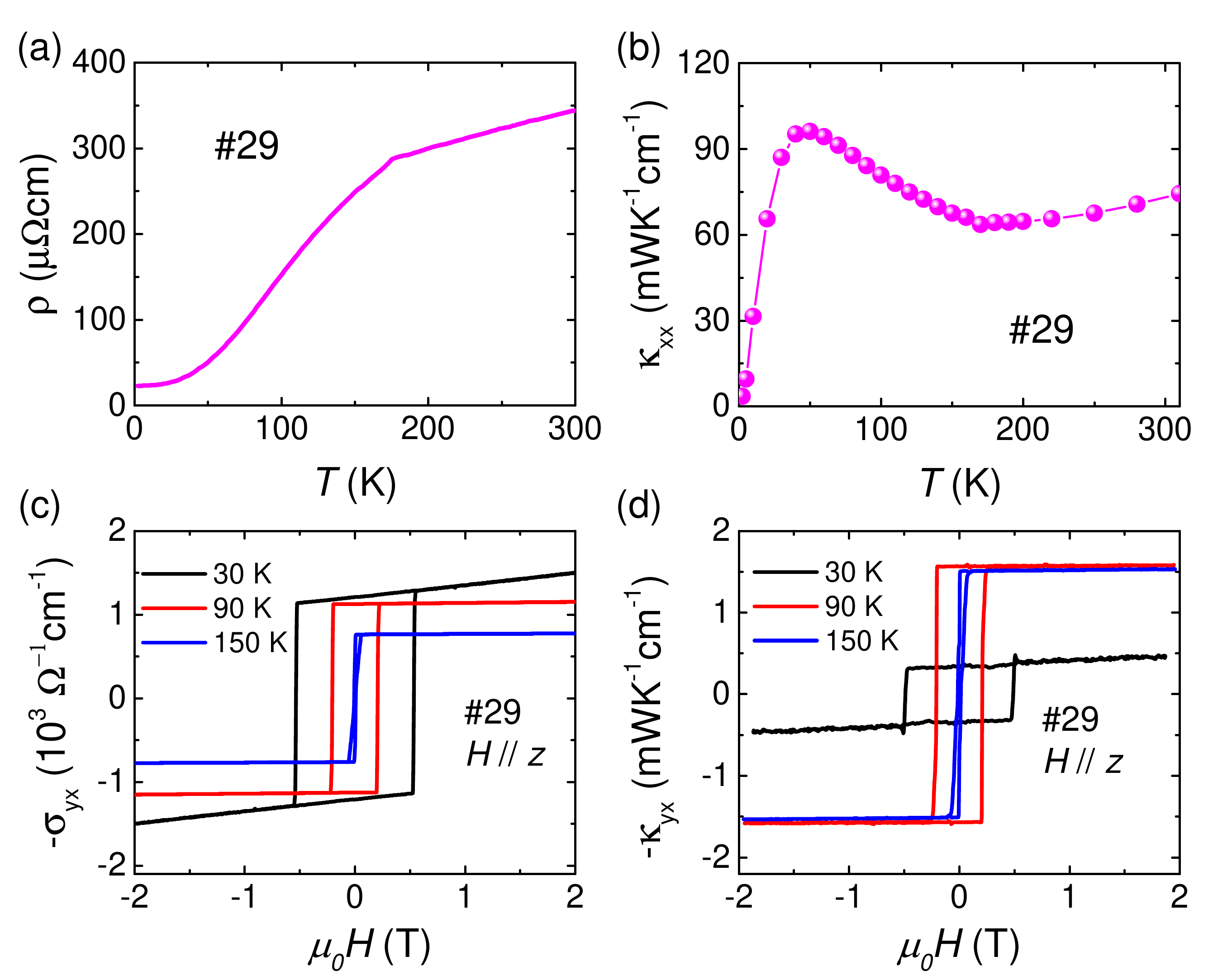}
		\caption{ (a) The temperature dependence of in-plane resistance and (b) thermal conductivity of the sample \#29.  The field dependence of (c) the Hall conductivity and (d) thermal Hall conductivity at various temperatures.}
		\label{SM.Fig.Hall}
	\end{figure}	
	
		\begin{figure}
		\includegraphics[width=9cm]{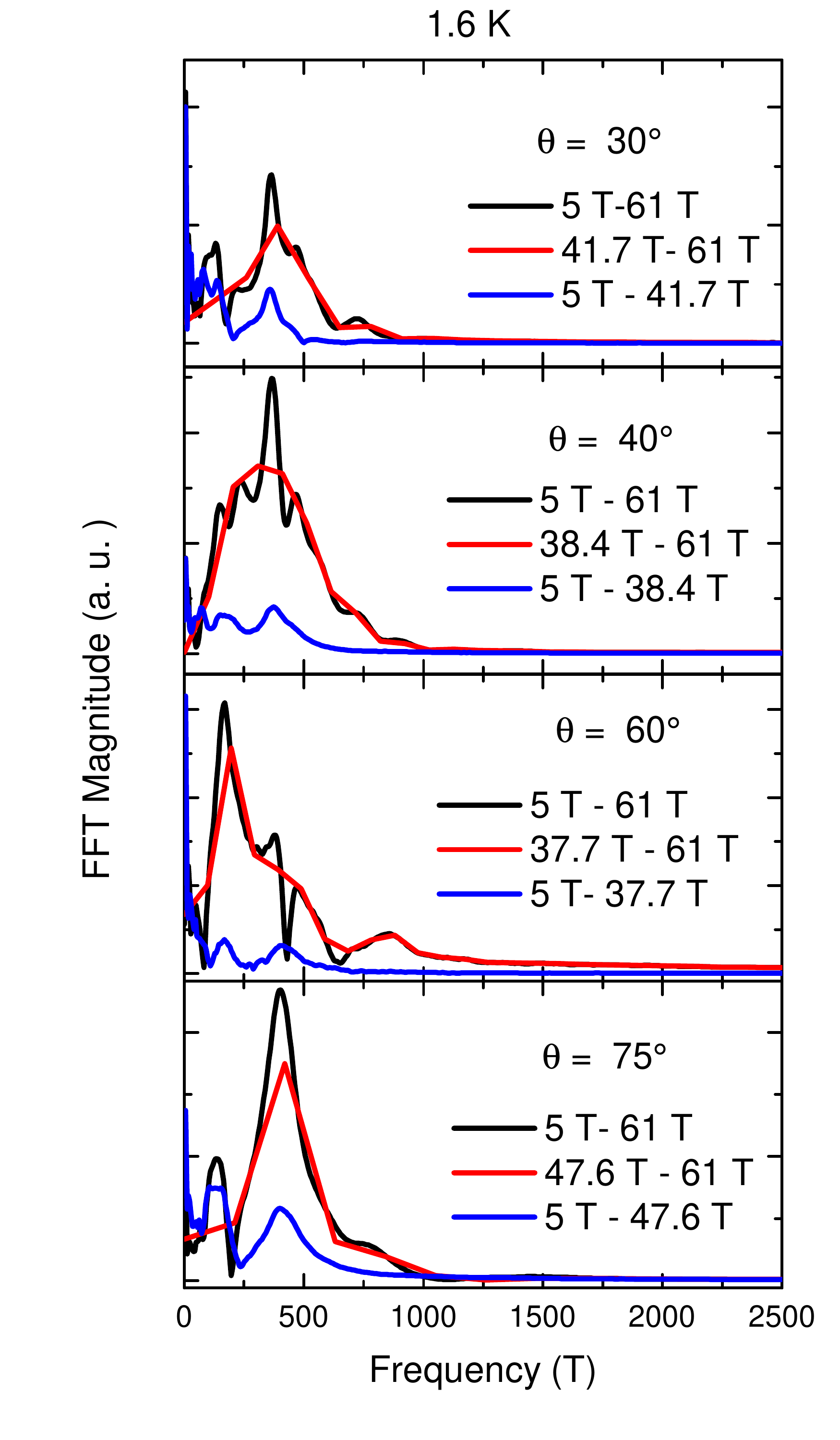}
		\caption{The breakdown fields for different angles were determined by carrying out FFT in different field ranges.}
		\label{SM.Fig.breakdownField}
	\end{figure}

\section{determining the in-plane orientation of the sample  }	

High quality single crystals of Co$_3$Sn$_2$S$_2$ were grown by the self-flux method, the nature cleavage plane is the $ab$-plane of the single crystals. The $c$-axis can be identified easily, while the orientations of the single crystal in $ab$-plane were confirmed by single-crystal XRD technique. The angle between $a$-axis and the $x$ is about 10$^\circ$, as the figure \ref{SM.xyz} shows.


\section{Raw data of Hall conductivity and thermal Hall conductivity}	
By measuring the field dependence of resistivity Fig. \ref{SM.Fig.Hall}(a) and Hall resistivity Fig. \ref{SM.Fig.Hall}(b), we can deduce the Hall conductivity through $\sigma_{yx}=\frac{-\rho_{yx}}{\rho_{xx}^2+\rho_{yx}^2}$ as the field is along $c$ or the $z$ axis. After plotted the Hall conductivity in the Fig. \ref{SM.Fig.Hall}(c), the anomalous Hall conductivity can be easily obtained by extrapolate the normal part of the Hall effect to zero field. The intersect at the the axis of the $\mu_0 H$=0 in the Fig. \ref{SM.Fig.Hall}(d) is the anomalous Hall conductivity. On the other hand, the thermal Hall conductivity can be obtained by measuring transverse thermal gradient with the measured longitudinal thermal conductivity $\kappa_{yx}=\frac{\partial_yT\kappa_{yy}\kappa_{xx}}{J_{qx}}$\cite{sm_Xiaokang2017}, where the $J_{qx}$ is the thermal density along $x$. The anomalous thermal Hall effect then is obtained by the same method as described for anomalous Hall effect.
		\begin{figure}
		\includegraphics[width=7cm]{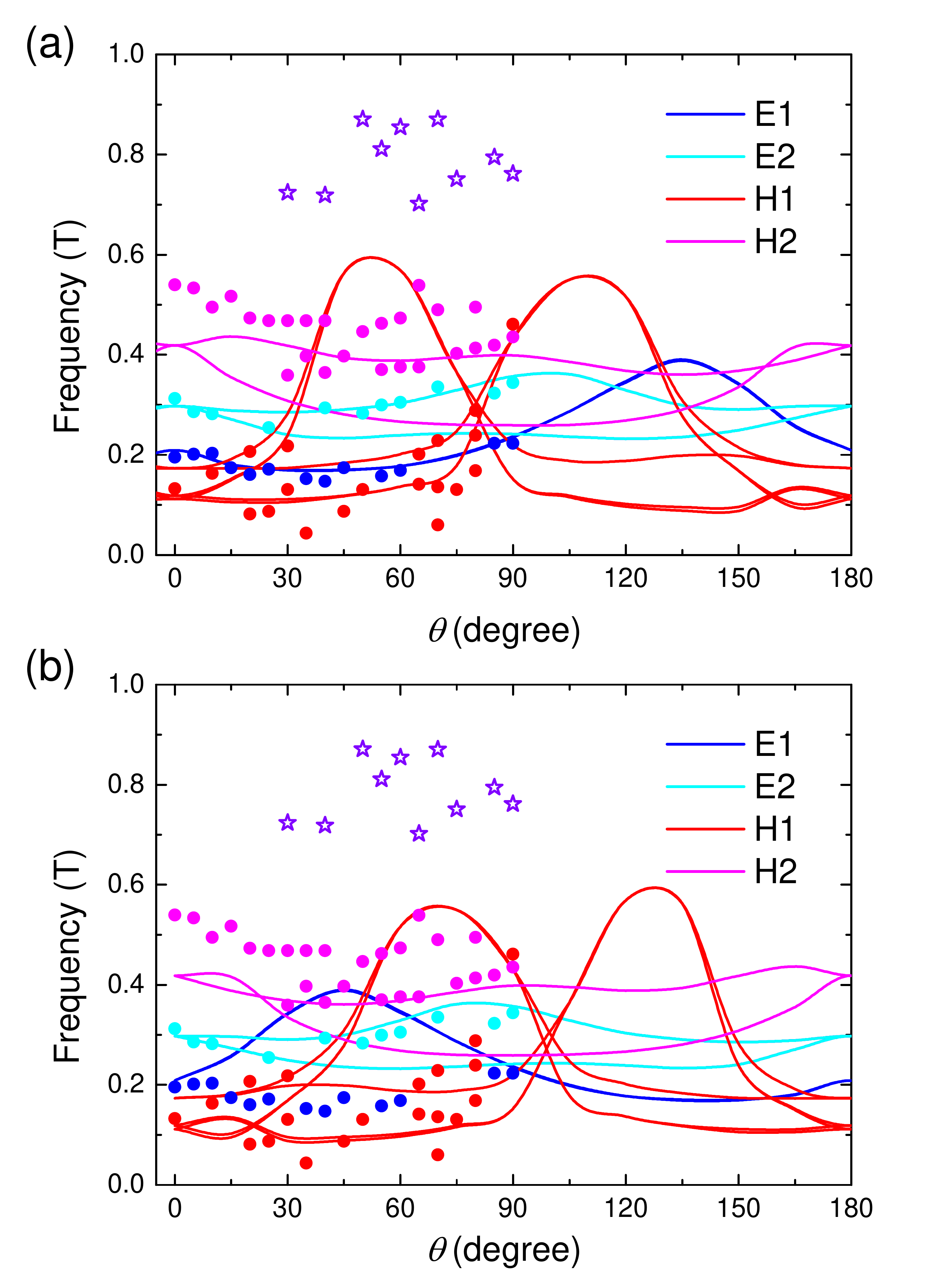}
		\caption{Angular dependence of the frequencies extracted from the SdH oscillations, the lines are from the theory calculation results for $z$ to $y$ direction (a) and --$z$ to $y$ direction(b). }
		\label{SM.Fig.Frequency}
	\end{figure}
	
\section{The breakdown fields for different angles}

		\begin{table}[!htb]
	\caption{The extracted frequency of the E2, H2 and the magnetic breakdowns at different angles, $H_B$ is where the magnetic breakdown field occurs.}
	\begin{tabular}{
			c|c| c| c| c}
		
		\hline
		\hline
		
		\rule{0pt}{10pt} $\theta$ & E2 (T) & H2 (T) & MB (T) &  $H_B (T)$\\
		\hline
		
		\rule{0pt}{9pt}	$90$$^{\circ}$  & 343 & 429 & 760 & 41.7 \\
		
		\rule{0pt}{9pt}	$85$$^{\circ}$  & 323 & 416 & 789 & 43.4\\
		\rule{0pt}{9pt}	$75$$^{\circ}$ &  & 403 & 751 & 47.6\\
		
		\rule{0pt}{9pt}	$65$$^{\circ}$ &  & 376 & 702 & 35.1 \\
		\rule{0pt}{9pt}	$60$$^{\circ}$ & 305 & 473 & 854 & 37.7 \\
		\rule{0pt}{9pt}	$55$$^{\circ}$  & 299 & 462 & 810 & 38.2 \\
		
		\rule{0pt}{9pt}	$40$$^{\circ}$  & 293 & 468 & 718 & 38.4 \\
       \rule{0pt}{9pt}	$30$$^{\circ}$  & & 468 & 724 & 41.7\\

		\hline
		\hline
	\end{tabular}
	
		\label{Table:breakdown}
	\end{table}

The breakdown fields for different angles were determined by carrying out FFT in different field ranges in the Fig. \ref{SM.Fig.breakdownField}. The black lines are for a field range from 5 T to 61 T. The blue lines are for the field range where no breakdown peak presents, while the red lines for the field range where breakdown peak presents. The table \ref{Table:breakdown} shows the summary of the results from experiments. The absence of the E2 in the  table \ref{Table:breakdown} is due to the small amplitude of the peak in the FFT. Below 30$^{\circ}$, the breakdown frequency is not observable.

\section{The angle-dependence of FFT lines in theoretical calculation }
	
As mentioned in the main text, the calculated SdH frequencies are not symmetric between +$z$ to $y$ direction (a) and --$z$ to $y$ direction(b), because of the low symmetry in the $yz$ plane. From our experimental data, we can't distinguish the plane the sample was rotated from +$z$ to $y$ direction or --$z$ to $y$ direction. By the comprising with the experimental data in the Fig.\ref{SM.Fig.Frequency}(a), the Fig. 3d shows better overlap between theoretical and experimental results.


\begin{thebibliography}{99}
		

\bibitem{PRL2014} Chen H, Niu Q and Macdonald A H 2014 Phys. Rev. Lett. \textbf{112} 017205
		
\bibitem {Nakatsuji2015} Nakatsuji S, Kiyohara N and Higo T, 2015 Nature \textbf{527} 212

\bibitem{Yan2017} Yan B  and Felser C 2017 Annu. Rev. Condens. Matter Phys. \textbf{8} 337

\bibitem{Smej} \v{S}mejkal L, Mokrousov Y, Yan B  and MacDonald A H  2018 Nat. Phys. \textbf{14} 242-251
\bibitem{Vaqueiro}	Vaqueiro P  and Sobany G G 2009 Solid State Sci. \textbf{11} 513

\bibitem{Holder} Holder M, Dedkov Y S, Kade A, Rosner H, Schnelle W, Leithe-Jasper A, Weihrich R and Molodtsov S L 2009 Phys. Rev. B \textbf{79} 205116

\bibitem{Schnelle} Schnelle W, Leithe-Jasper A, Rosner H, Schappacher F M,
P\"{o}ttgen R, Pielnhofer F and Weihrich R 2013 Phys. Rev. B \textbf{88}, 144404

\bibitem {Behnia2016} Behnia K and Aubin H  2016 Rep. Prog. Phys. \textbf{79} 046502

\bibitem {Ding2019} Ding L, Koo J, Xu L, Li X, Lu X, Zhao L, Wang Q, Yin Q, Lei H, Yan B, Zhu Z and Behnia K 2019 Phys. Rev. X \textbf{91} 41061
		
		
		
		
		
		
		
		
		
		
		
	
\bibitem{LiuEnke2018} Liu E, Sun Y, Kumar N, Muechler L, Sun A, Jiao L,  Yang S-Y, Liu D, Liang A, Xu Q, Kroder J, S\"{u}\ss V, Borrmann H, Shekhar C, Wang Z, Xi C, Wang W, Schnelle W, Wirth S, Chen Y, Goennenwein S T B and Felser C 2018 Nat. Phys. \textbf{14} 1125
		
\bibitem{Lei2018} Wang Q, Xu Y, Lou R, Liu Z, Li M, Huang Y, Shen D, Weng H, Wang S and Lei H 2018 Nat. Commun. \textbf{9} 3681
		
		
\bibitem{AFM2020} J Shen, Zeng Q, Zhang S, Sun H, Yao Q, Xi X, Wang W, Wu G, Shen B, Liu Q  and Liu E 2020 Adv. Funct. Mater. \textbf{32} 2000830
		
\bibitem{ChenYulin2019} Liu D F, Liang A J, Liu E K, Xu Q N, Li Y W, Chen C, Pei D, W. Shi J, Mo S K, Dudin P, Kim T, Cacho C, Li G, Sun Y, Yang L X, Liu Z K, Parkin S S P, Felser C and Chen Y L 2019 Science \textbf{365} 1282

\bibitem{Morali2019} Morali N, Batabyal R, Nag P K, Liu E, Xu Q, Sun Y, Yan B, Felser C, Avraham N and Beidenkopf H 2019 Science \textbf{365}  1286
		
		
\bibitem{MO2020} Okamura Y, Minami S, Kato Y, Fujishiro Y, Kaneko Y, Ikeda J, Muramoto J, Kaneko R, Ueda K, Kocsis V, Kanazawa N, Taguchi Y, Koretsune T, Fujiwara K, Tsukazaki A, Arita R, Tokura Y and Takahashi Y 2020 Nat. Commun. \textbf{11} 4619
	

\bibitem{Brien2016}O'Brien T E, Diez M and Beenakker C W J 2016 Phys. Rev. Lett. \textbf{116} 236401
		
\bibitem {Zhu2015} Zhu Z, Lin X, Liu J, Fauqu\'{e} B, Tao Q, Yang C, Shi Y and Behnia K 2015 Phys. Rev. Lett. \textbf{114} 176601

\bibitem {Shoenberg} D. Shoenberg, Magnetic oscillations in metals, Cambridge University Press (1984)

\bibitem{Muller2020} M\"{u}ller C S A, Khouri T, van Delft M R, Pezzini S, Hsu Y T, Ayres J, Breitkreiz M, Schoop L M, Carrington A, Hussey N E; Wiedmann S 2020 Physical Review Research \textbf{2} 023217


\bibitem{Xu2020} Xu L, Li X, Lu X, Collignon C, Fu H, Koo J,Fauqu\'{e} B, Yan B, Zhu Z and Behnia K 2020 Sci. Adv. \textbf{6} eaaz3522

\bibitem{Li2018} Li X, Xu L, Ding L, Wang J, Shen M, Lu X, Zhu Z and Behnia K 2017  Phys. Rev. Lett. \textbf{119} 056601


\bibitem {Growth2020} Xu Y, Zhao J, Yi C, Wang Q, Yin Q, Wang Y, Hu X, Wang L, Liu E, Xu G, Lu L, Soluyanov A A, Lei H, Shi Y, Luo J and Chen Z G 2020 Nat. Commun. \textbf{11} 3985	

\bibitem{PBE1996} Perdew J P, Burke K and Ernzerhof M 1996 Phys. Rev. Lett. \textbf{77} 3865

\bibitem{Kresse1999} Kresse G  and Joubert D 1999 Phys. Rev. B \textbf{59} 1758

\bibitem{Mostofi2008} Mostofi A A, Yates J R, Pizzi G, Lee Y S, Souza I, Vanderbilt D and Marzari N 2014 Comput. Phys. Commun. \textbf{185} 2309-2310

\bibitem{SM} Please see the SM for more discussions

\bibitem{Tanaka2020} Tanaka M, Fujishiro Y, Mogi M, Kaneko Y, Yokosawa T, Kanazawa N, Minami S, Koretsune T, Arita R, Tarucha S, Yamamoto M and Tokura Y 2020 Nano Lett. \textbf{20}  7476-7481
		
\bibitem{nano2020} Yang S Y, Noky J, Gayles J, Dejene F K, Sun Y, D\"{o}rr M, Skourski Y, Felser C, Ali M N, Liu E and Parkin S S P 2020  Nano Lett. \textbf{20} 7860-7867

\bibitem{Ali2014}  Ali M N, Xiong J, Flynn S, Tao J, Gibson Q D, Schoop L M, Liang T, Haldolaarachchige N, Hirschberger M, Ong N P and Cava R J 2014 Nature \textbf{514} 205-208
		
		
\bibitem {Kumar2017} Kumar N, Sun Y, Xu N, Manna K, Yao M, S\"{u}ss V, Leermakers I, Young O, F\"{o}rster T, Schmidt M, Borrmann H, Yan B, Zeitler U, Shi M, Felser C and Shekhar C 2017 Nat. Commun. \textbf{8} 1642
		
\bibitem {Gao2017} Gao W, Hao N, Zheng F W, Ning W, Wu M, Zhu X, Zheng G, Zhang J, Lu J, Zhang H, Xi C, Yang J, Du H, Zhang P, Zhang Y and Tian M 2017 Phys. Rev. Lett. \textbf{118} 256601
		
\bibitem {Zhao2018} Zhao L, Xu L, Zuo H, Wu X, Gao G and Zhu Z 2018 Phys. Rev. B \textbf{98} 085137

\bibitem {Fauque2018}  Fauqu\'e B, Yang X, Tabis W, Shen M, Zhu Z, Proust C, Fuseya Y, and Behnia
K 2018 Phys. Rev. Materials \textbf{2} 114201

\bibitem {Shen2019} Shen J, Zeng Q, Zhang S, Tong W, Ling L, Xi C, Wang Z, Liu E, Wang W, Wu G and Shen B 2019 Appl. Phys. Lett. \textbf{115} 212403
\bibitem{YBCO2009}  Audouard A, Jaudet C, Vignolles D, Liang R, Bonn D A, Hardy W N, Taillefer L and Proust C 2009 Phys. Rev. Lett. \textbf{103}  157003

\bibitem {LK1956} Lifshits I and Kosevich A M 1956 Sov. Phys. JETP \textbf{2}  636-645

\bibitem{Jaoui2020} Jaoui A, Fauqu\'{e} B and Behnia K 2020 Nat. Commun. \textbf{12} 195



\bibitem{Yang2020}Yang R, Zhang T, Zhou L, Dai Y, Liao Z, Weng H and Qiu X  2020 Physical Review Letters \textbf{124} 077403
	
	

		
\bibitem{Xu2020b}	Xu L, Li X, Ding L, Chen T, Sakai A, Fauqu\'{e} B, Nakatsuji S , Zhu Z  and Behnia K 2020 Phys. Rev. B \textbf{101} 180404(R)	


\bibitem{Haldane} Haldane F D M 2004 Phys. Rev. Lett. \textbf{93} 206602		
		
		
		

		


		

	\end{thebibliography}

\begin{thebibliography}{99}

		

	\bibitem {sm_Xiaokang2017} Li X, Xu L, Ding L, Wang J, Shen M, Lu X, Zhu Z and Behnia K 2017 Phys. Rev. Lett. \textbf{119} 056601
	
	
	
	\end{thebibliography}
\end{document}